\documentclass[12pt,preprint]{aastex}
\usepackage{rotating}
\usepackage{graphicx}
\usepackage{color}

\slugcomment{The Astronomical Journal}

\shorttitle{313P/Gibbs}
\shortauthors{Jewitt et al.}

\begin{document}

\title{New Active Asteroid 313P/Gibbs}
\author{David Jewitt$^{1,2}$,  Jessica Agarwal$^3$, Nuno Peixinho$^{4}$, Harold Weaver$^5$,  Max Mutchler$^6$,  Man-To Hui$^{1}$, Jing Li$^{1}$, and Stephen Larson$^7$
}
\affil{$^1$Department of Earth, Planetary and Space Sciences,
UCLA, 
595 Charles Young Drive East, 
Los Angeles, CA 90095-1567\\
$^2$Dept.~of Physics and Astronomy,
University of California at Los Angeles, \\
430 Portola Plaza, Box 951547,
Los Angeles, CA 90095-1547\\
$^3$ Max Planck Institute for Solar System Research, Max-Planck-Str. 2, 37191 Katlenburg-Lindau, Germany\\
$^4$ Unidad de Astronom\'{\i}a, Fac.~de Ciencias B\'asicas, Universidad de Antofagasta, Avda.~U.~de Antofagasta 02800, Antofagasta, Chile\\
$^5$ The Johns Hopkins University Applied Physics Laboratory, 11100 Johns Hopkins Road, Laurel, Maryland 20723  \\
$^6$ Space Telescope Science Institute, 3700 San Martin Drive, Baltimore, MD 21218 \\
$^7$ Lunar and Planetary Laboratory, University of Arizona, 1629 E. University Blvd.
Tucson AZ 85721-0092 \\
}

\email{jewitt@ucla.edu}

\begin{abstract}
We present initial observations of the newly-discovered active asteroid 313P/Gibbs (formerly P/2014 S4), taken to characterize its nucleus and  comet-like activity.  The central object has a radius $\sim$0.5 km (geometric albedo 0.05 assumed). We find no evidence for secondary nuclei and set (with qualifications) an upper limit to the radii of such objects near 25 m, assuming the same albedo.   Both aperture photometry and a morphological analysis of the ejected dust show that  mass-loss is continuous at rates $\sim$0.2 to 0.4 kg s$^{-1}$, inconsistent with an impact origin.  Large dust particles, with radii $\sim$50 to 100 $\mu$m, dominate the optical appearance. At 2.4 AU from the Sun, the surface equilibrium temperatures are too low for thermal or desiccation stresses to be responsible for the ejection of dust.  No gas is spectroscopically detected (limiting the gas mass loss rate to $<$1.8 kg s$^{-1}$). However, the protracted emission of dust seen in our data and the detection of another episode of dust release near perihelion, in archival observations from 2003, are highly suggestive of an origin by the sublimation of ice.  Coincidentally, the  orbit of 313P/Gibbs is similar to those of several active asteroids  independently suspected to be  ice sublimators, including  P/2012 T1, 238P/Read and 133P/Elst-Pizarro, suggesting that ice is abundant in the outer asteroid belt.
\end{abstract}


\section{INTRODUCTION}
\begin{figure}
\epsscale{0.85}
\begin{center}
\plotone{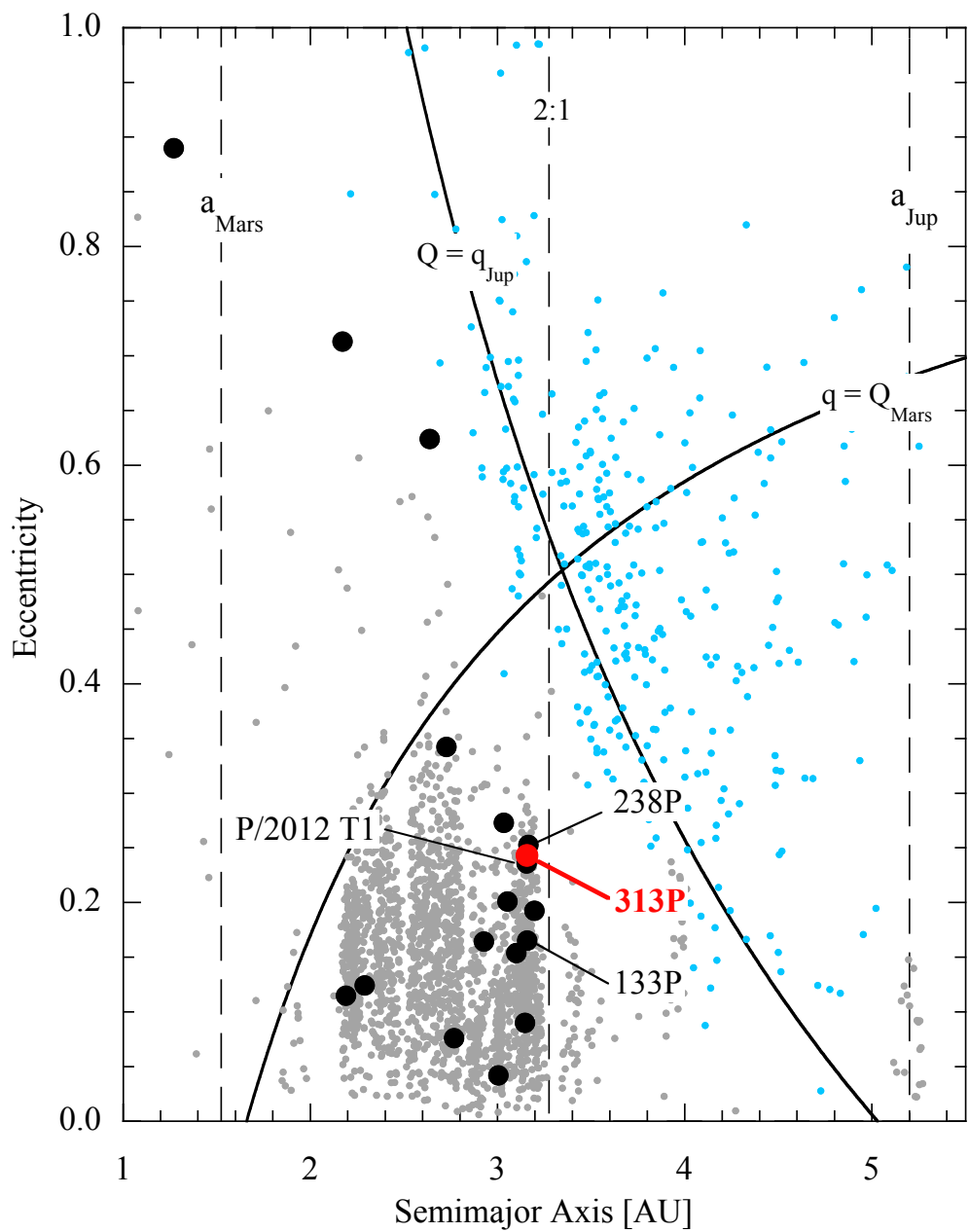}
\caption{Orbital semimajor axis vs.~eccentricity for asteroids (grey circles), comets (blue circles) and active asteroids (large black circles).  313P is distinguished by a red circle.  Vertical dashed lines mark the semimajor axes of Mars and Jupiter and the location of the 2:1 mean-motion resonance with Jupiter.  Objects plotted above the diagonal arcs have either a perihelion distance smaller than the aphelion distance of Mars, or an aphelion distance larger than Jupiter's perihelion distance, and are consequently dynamically short-lived.    \label{ae_plot}
} 
\end{center} 
\end{figure}

313P/Gibbs (formerly P/2014 S4), hereafter called ``313P'', was discovered on UT 2014 September 24 as a product of the on-going Catalina Sky Survey (Gibbs 2014). Pre-discovery observations  allowed the orbit to be accurately established, with semimajor axis 3.156 AU, eccentricity 0.242 and inclination 11.0\degr.  The Tisserand parameter with respect to Jupiter is $T_J$ = 3.13, consistent with main-belt membership and distinct from the orbits of Kuiper belt and Oort cloud comets, which have $T_J <$ 3 (Kresak 1982).  This is a new member of the active asteroids (a.k.a.~main-belt comet) population (Hsieh and Jewitt 2006), with an orbit in the outer asteroid belt  (Figure \ref{ae_plot}).  

Dynamical transfer to the asteroid belt from the Kuiper belt and Oort cloud comet reservoirs  is highly inefficient in the modern solar system, especially for low-inclination orbits  (Fernandez et al.~2002, Levison et al.~2006). Indeed, long-term integrations of several active asteroids have shown billion-year stability of their orbits (Haghighipour 2009), and an in-situ origin seems likely for most.  The active asteroids are driven by a surprising variety of mechanisms, from impact, to sublimation, to thermal fracture and to rotational instabilities, all previously thought to lie beyond the realm of observation (Jewitt 2012).    

We initiated a program of observations designed to characterize 313P and to determine the origin of its activity; here we present its initial results.

\section{OBSERVATIONS} 

Observations described here were taken using three telescopes.  The Danish 1.5 meter telescope employed the DFOSC camera, which has a 0.39\arcsec~pixel$^{-1}$ image scale and a 13.7\arcmin$\times$13.7\arcmin~field of view (Andersen et al.~1995).  Typical image quality was $\sim$1.4\arcsec.  The observations were taken through a broadband R filter with the telescope tracked sidereally and short (180 s) exposures used to minimize trailing.  The Hubble Space Telescope (HST) was used with the WFC3 camera (Dressel 2012) whose 0.04\arcsec~pixels each correspond to about 42 km  at the $\sim$1.4 AU distance of 313P. The Nyquist-sampled spatial resolution is $\sim$84 km.  All observations were taken using the very broad F350LP filter (4758\AA~FWHM) which has an effective wavelength of 6230\AA~on a solar-type (G2V) source.  From each orbit we obtained five exposures of 410 s duration.  The Keck 10-m telescope was used   to image and to obtain  spectra of 313P in search of emission lines from gas.  We employed the two-channel LRIS camera (Oke et al.~1995) having imaging scale 0.135\arcsec~pixel$^{-1}$ and tracked the telescope non-sidereally while autoguiding on fixed stars.  Typical image quality during the observations was 0.8 to 1.0\arcsec~FWHM and the nights were photometric to better than $\pm$2\%.   We used broadband Kron-Cousins B, V and R filters, with calibration from nearby Landolt stars.

The observational geometry for the observations is summarized in Table \ref{geometry}.

 \section{RESULTS} 
 
 \subsection{Morphology}
\begin{figure}
\epsscale{0.9}
\begin{center}
\plotone{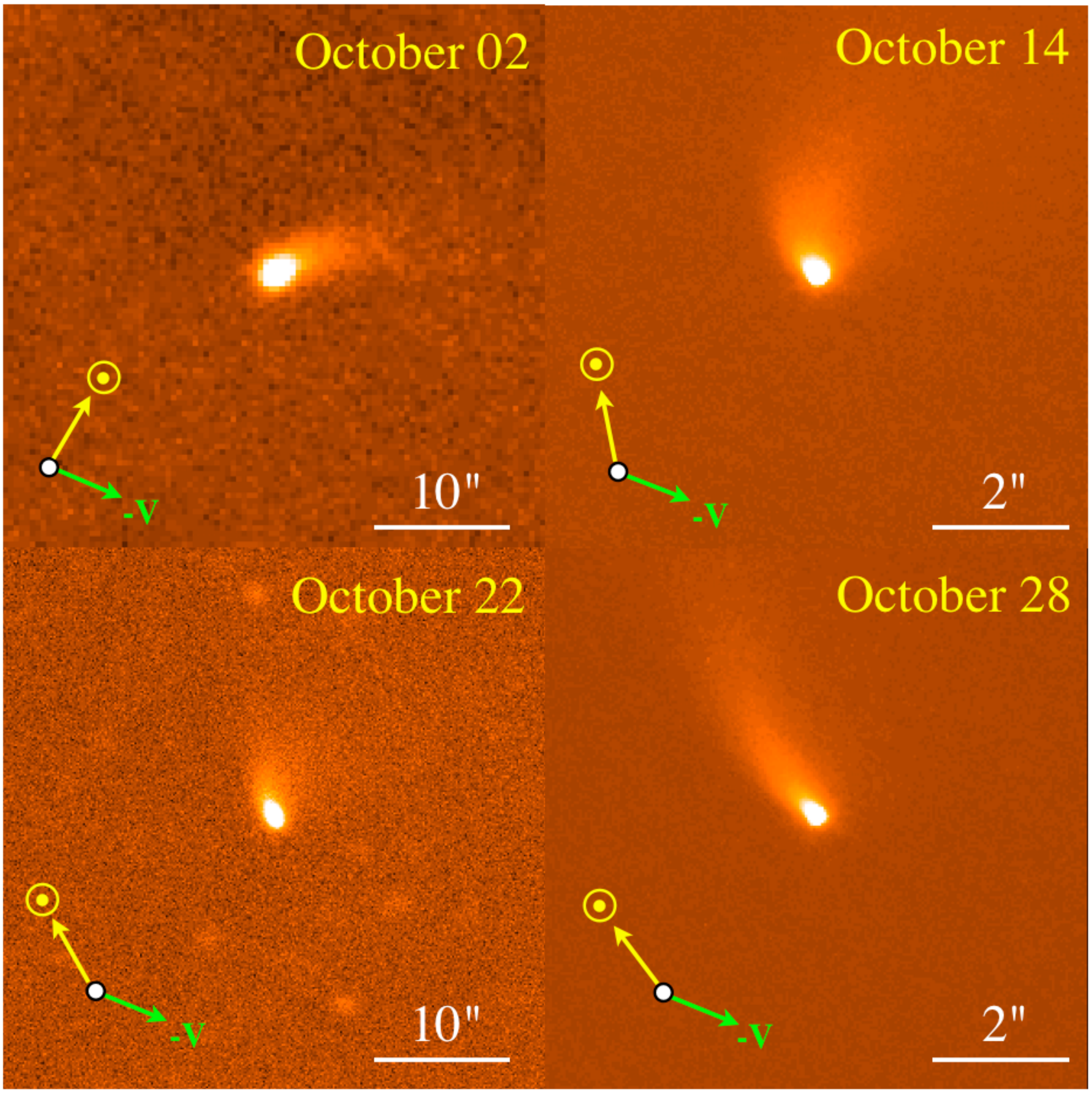}
\caption{Active asteroid 313P imaged on four UT dates in 2014, as marked.  The yellow arrow in each panel shows the direction to the projected antisolar vector while the green arrow indicates the projected anti-velocity vector (see Table \ref{geometry}).  Each panel has North to the top, East to the left. \label{images}
} 
\end{center} 
\end{figure}

Sample images of 313P are shown in Figure (\ref{images}) along with vectors to indicate the anti-solar direction (yellow) and the negative of the heliocentric velocity (green), projected into the plane of the sky.  Images from adjacent dates (October 02 and 03 (Danish telescope) and from October 22 and 23 (Keck)) are visually similar and so are not shown separately in the figure, while images from November 06 are strongly affected by scattered moonlight and were used only to obtain an integrated magnitude.  The object shows a single point-like nucleus from which emerges a diffuse, fan-shaped tail having clockwise curvature.  The anti-solar direction undergoes a large rotation as a result of changes in the observing geometry near opposition (Table \ref{geometry}).   The fan-shaped tail approximately follows this rotation while preserving its curvature in each of the panels of the figure.  The morphology clearly shows that this is a dust tail.

\begin{figure}
\centering
\epsscale{0.65}
\begin{center}
\includegraphics[width=0.90\textwidth,angle=0]{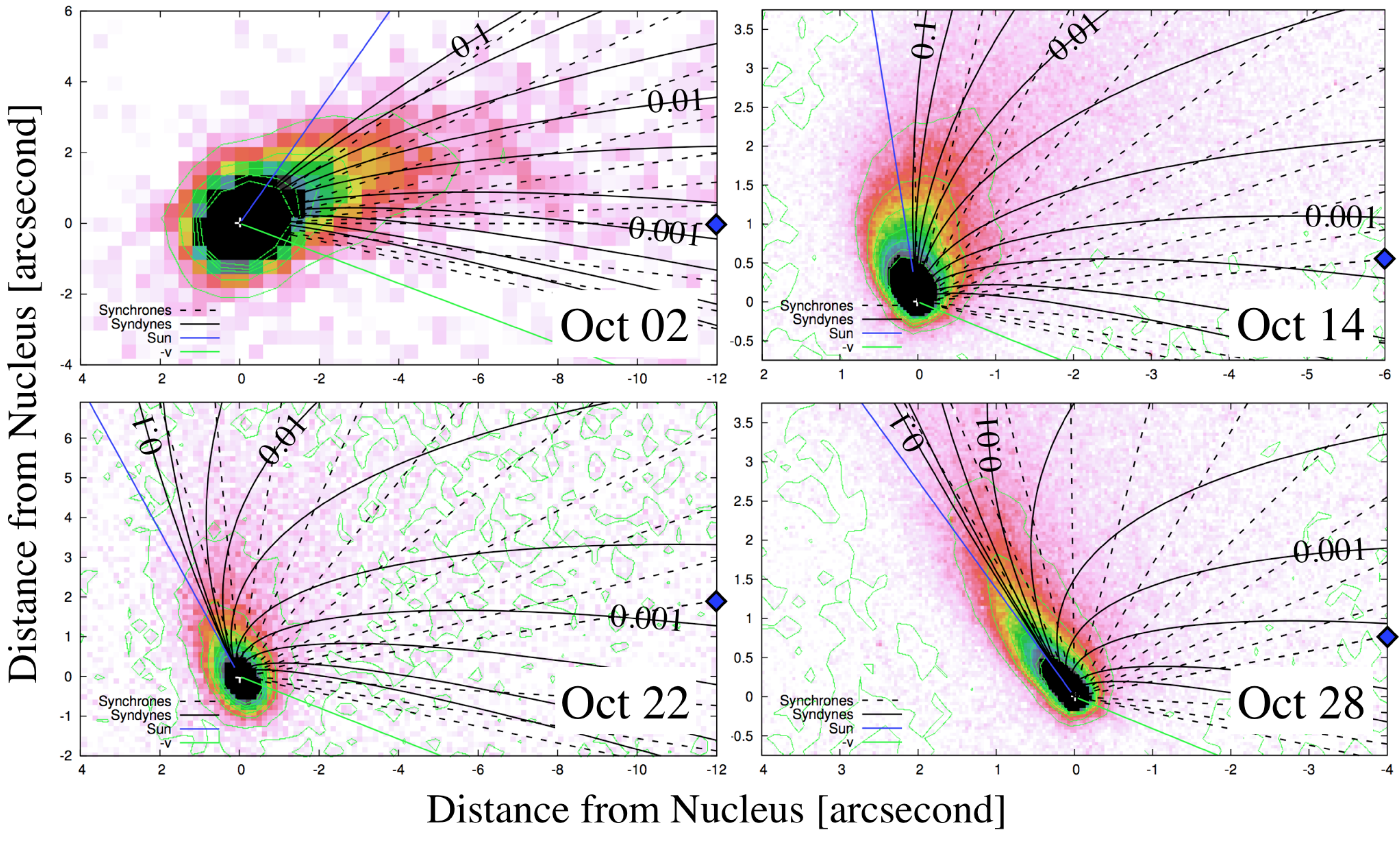}
\caption{
Syndynes (solid black lines) and synchrones (dashed black lines) superimposed on images from four dates in 2014.   Syndynes are plotted for $\beta$ = 0.1, 0.05, 0.02, 0.01, 0.005, 0.002, 0.001, 0.0005, 0.0002, 0.0001.  Only three syndynes are labeled for clarity.  Synchrones dates increase anti-clockwise with intervals of 50 days before July-9 and 10 days afterwards.  Only the July 9 synchrone is marked (by a blue diamond) to avoid clutter on the plots.  The directions of the projected antisolar vector (blue line) and negative heliocentric velocity vector (green line) are shown.  Each image has north to the top and east to the left.  Distances are offsets from the nucleus in arcseconds.
\label{models}
} 
\end{center} 
\end{figure}

The trajectories of dust particles relative to their parent nucleus are determined by the ratio of  radiation pressure acceleration to local solar gravity, $\beta$, and by the initial velocity of ejection.  The $\beta$ ratio is inversely proportional to particle radius, meaning that the paths of ejected dust grains can be used to estimate the dust size from $\beta$ (Bohren and Huffman 1983).  Figure (\ref{models}) shows a set of syndynes (loci of positions of particles of a given $\beta$ released at different times) and synchrones (positions of particles having a range of $\beta$ but released from the nucleus at one time) over-plotted on data from four dates in 2014 October.  The syndyne and synchrone models are computed under the assumption that the dust particles leave the nucleus with zero initial velocity (Finson and Probstein 1968).  Syndynes and synchrones represent two extremes of the style of particle ejection, and there is no physical reason to expect that a given comet should match one extreme or the other.  However, Figure (\ref{models}) clearly shows that the tail of 313P is better approximated by syndyne trajectories than by synchrones.  The syndynes, for instance, naturally reproduce the clockwise curvature of the tail seen in the imaging data.  We therefore conclude that mass loss was more nearly continuous over the period of observation than impulsive.  This is different, for example, from active asteroid 311P/PanStarrs in which a multiple tail system is very well matched by a set of synchrones (Jewitt et al.~2013).  The axis of the dust tail is well-matched by particles with $\beta \sim$ 0.01 to 0.02, corresponding to grain radii $a \sim$ 50 to 100 $\mu$m.   

A strong upper limit to the sunward extent of the coma of 313P may be set at 0.8\arcsec, corresponding to distance $\ell$ = 900 km in the plane of the sky.  This is small compared to the sunward extents of the comae of active Jupiter family and Oort cloud comets  and implies a low sunward dust ejection velocity.  If we take $\ell$ as the turn-around distance of a dust particle ejected towards the Sun, and assume that the acceleration of the particle is constant, then the ejection speed is simply $V = (2 \beta g_{\odot} \ell)^{1/2}$.  At heliocentric distance $R$ = 2.4 AU, the local solar gravity is $g_{\odot}$ = 10$^{-3}$ m s$^{-2}$.  Substituting for $\ell$ and $\beta$ gives $V \ll$ 4 to 6 m s$^{-1}$.  This estimate is very crude because it neglects the effects of projection (so that the sky-plane $\ell$ is a lower limit to the true $\ell$) but nevertheless indicates that the ejection of optically dominant dust particles in 313P is slow.  Very low ejection speeds were also measured in probable ice-sublimating active asteroid 133P/Elst-Pizarro and attributed to a small gas-drag acceleration length resulting from the small physical size of the exposed ice patch (Jewitt et al.~2014b). The angular extents of the comae of Jupiter family and Oort cloud comets are typically comparable to an arcminute (i.e.~about 10$^2$ times larger than in 313P) at the same heliocentric distance, and their characteristic dust radiation pressure factors are 50 to 100 times larger than in 313P (c.f.~Jewitt and Meech 1987).  As a result, the ejection speeds implied by the above relation are about 10$^2$ times larger than in 313P and comparable to the thermal speeds in sublimated gas.  The latter, given a local blackbody temperature of 180 K at 2.4 AU, is  about 450 m s$^{-1}$.  

\subsection{Photometry}
\label{dust}

The apparent magnitudes of 313P, $m_{\lambda}$ (where $\lambda$ is the effective wavelength of the filter used for the observation), were measured using circular projected photometry apertures up to 6.0\arcsec~in radius, with background subtraction from a contiguous annulus having 12.0\arcsec~outer radius (Table \ref{photometry}).   We used small (0.2\arcsec~radius) aperture photometry from HST to constrain the nucleus of 313P and larger apertures to measure primarily scattering of sunlight by  ejected dust.  Photometric calibration of the ground-based data was obtained through near-simultaneous measurements of standard stars from the list by Landolt (1992).  Calibration of the HST data was obtained using the HST exposure time calculator, from which we found that a V = 0 G2V star gives a count rate 4.72$\times$10$^{10}$ s$^{-1}$ when measured with a 0.2\arcsec~radius aperture.   Uncertainties in Table (\ref{photometry}) were estimated from scatter in repeated measurements and do not include potential systematic errors caused, for example, by small differences in the filter systems used on different telescopes.  We estimate that these errors are comparable to the statistical uncertainties listed in the Table, except that systematic errors on the HST data are larger ($\pm$10\%) because of the use of the very broad F350LP filter and the uncertain transformation to standard astronomical V magnitudes.

Keck photometry summarized in Table (\ref{photometry}) indicates colors B-V = 0.72$\pm$0.02, V-R = 0.36$\pm$0.02, B-R = 1.08$\pm$0.02 on UT 2014 October 22 and B-V = 0.72$\pm$0.02, V-R = 0.40$\pm$0.02, B-R = 1.12$\pm$0.02 on UT 2014 October 23  within a circular photometry aperture of projected radius 6.0\arcsec.  For comparison, the solar colors are  $(B-V)_{\odot}$ = 0.64$\pm$0.02, $(V-R)_{\odot}$ = 0.35$\pm$0.02 and $(B-R)_{\odot}$ = 0.99$\pm$0.02 (Holmberg et al.~2006). Thus, 313P is slightly redder than the Sun, consistent with a normalized continuum reflectivity gradient $S' = 5 \pm 2$ \%/1000\AA~across the optical spectrum and suggestive of the optical colors of C-type asteroids, which are abundant in the outer asteroid belt (DeMeo and Carry 2013).  As a note of caution, however, it should be remarked that these colors refer to the dust which dominates the scattering cross-section of 313P and that the nucleus could have, in principle, a different color.  

The apparent magnitudes were corrected to unit heliocentric, $R$, and geocentric, $\Delta$, distance and to zero phase angle, $\alpha$, using the inverse square law

\begin{equation}
m_{\lambda}(1,1,0) = m_{\lambda} - 5\log(R\Delta) + 2.5\log_{10}(\Phi(\alpha)).
\label{absolute}
\end{equation}

\noindent Here, $\Phi(\alpha)$ is the phase function at phase angle $\alpha$, equal to the ratio of the scattered light at $\alpha$ to that at $\alpha$ = 0\degr.   We assumed the phase function formalism of Bowell et al.~(1989) with parameter $g$ = 0.15, as appropriate for a C-type object.  The phase function of 313P is unmeasured, introducing a small uncertainty into the value of $m_{\lambda}(1,1,0)$.   However, even at the largest phase angles of the present observations ($\alpha$ = 13\degr, Table \ref{geometry}), the difference between assumed C-type and S-type phase corrections is only 0.1 magnitudes, giving an estimate of the magnitude of the phase correction uncertainty.   Absolute magnitudes are given in Table (\ref{photometry}) with their statistical uncertainties.

The absolute magnitudes are related to the effective scattering cross-section of the material within the photometry aperture, $C_e$ (km$^2$),  by

\begin{equation}
C_e = \frac{2.24\times10^{16} \pi}{p_{\lambda}} ~10^{-0.4[m_{\odot, \lambda} - m_{\lambda}(1,1,0)]}
\label{area}
\end{equation}

\noindent where $p_{\lambda}$ is the geometric albedo of 313P and $m_{\odot, \lambda}$ is the apparent magnitude of the Sun, both at wavelength $\lambda$.  We assume $V_{\odot}$ = -26.77.  The resulting scattering cross-sections are listed in Table (\ref{photometry}), computed assuming $p_{V}$ = 0.05 with small adjustments for $p_B$ and $p_R$ as indicated by the broadband colors.  Uncertainties on $C_e$ again reflect only statistical uncertainties, not systematic errors introduced by the phase function or uncertainties in the magnitude and color of the Sun.

\begin{figure}
\epsscale{0.95}
\begin{center}
\includegraphics[width=0.95\textwidth,angle=0]{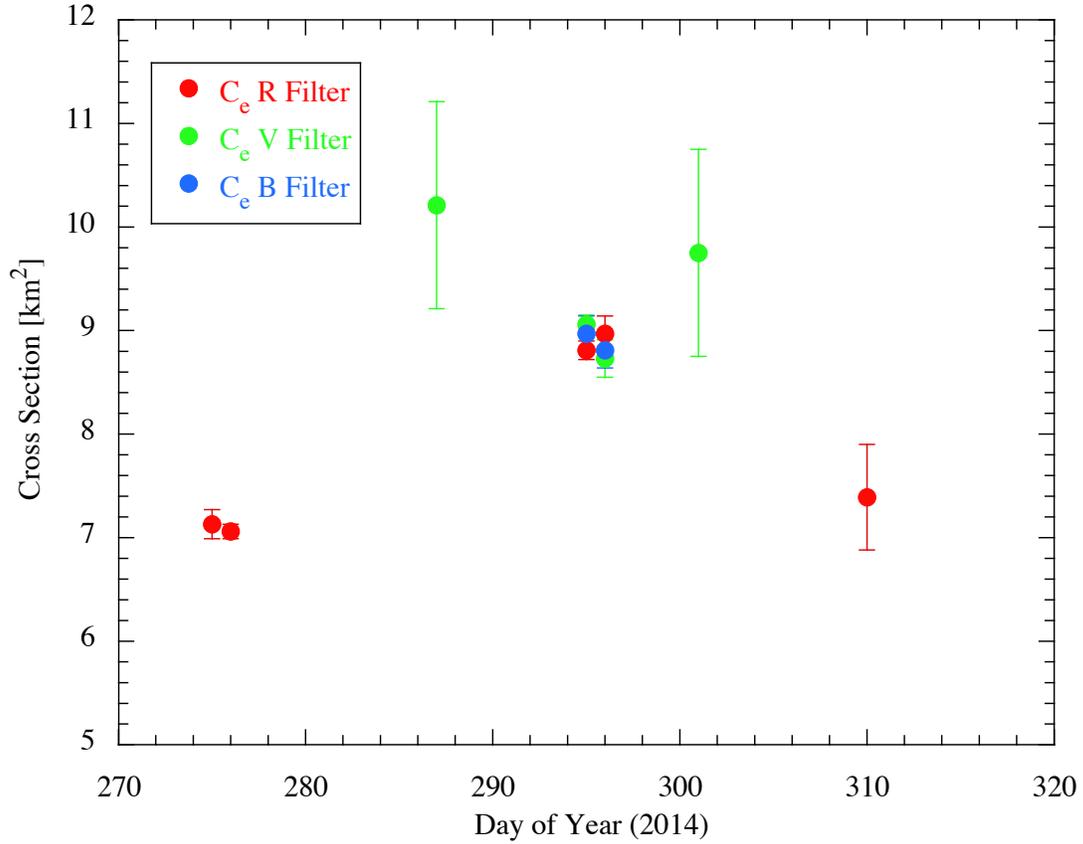}
\caption{Scattering cross-section (km$^2$) vs.~day of year, data from Table (\ref{photometry}).  The measurements at DOY 287 and 301 have been plotted with $\pm$10\% error bars to illustrate the systematic uncertainty resulting from the use of the very broad F350LP filter and the uncertainty in the transformation to standard V magnitudes.  The measurement at DOY 310 has a larger uncertainty owing to scattered moonlight. \label{Ce_vs_time}
} 
\end{center} 
\end{figure}

Figure (\ref{Ce_vs_time}) shows the scattering cross-section, $C_e$, derived from 6.0\arcsec~radius aperture photometry (Table \ref{photometry}), plotted as a function of time (expressed as Day of Year (DOY), c.f.~Table \ref{geometry}).  The Figure shows that the cross-section rises over a $\sim$month-long period, with the peak value, $C_e$ = 10 km$^2$ (DOY 287) occurring $\Delta t$ = 12 days after the minimum $C_e$ = 7 km$^2$ (DOY 275).   The increase in the cross-section by $\Delta C_e$ = 3 km$^2$ can be interpreted in terms of the ejected dust mass using $\Delta M_d = 4/3\rho \overline{a} \Delta C_e$, where $\rho$ is the dust density and $\overline{a}$ is the mean particle radius.  We assume a nominal $\rho$ = 10$^3$ kg m$^{-3}$ and set $\overline{a}$ = 50 to 100 $\mu$m (as suggested by the syndyne fits in Figure \ref{models}), to find $\Delta M_d \sim$ (2 to 4)$\times$10$^5$ kg.  The corresponding effective mass loss rate is $dM_d/dt \sim \Delta M_d/\Delta t \sim$ 0.2 to 0.4 kg s$^{-1}$.  Obviously, this  estimate is very crude and can be improved by better treatment of the dust particle size distribution (to better define $\overline{a}$) and by the acquisition of photometry with a higher time resolution.  Strictly, this is the increase in the mass loss rate over and above the rate on DOY 275.  Nevertheless, it serves to show that the tail of 313P can be supplied by a very modest ($<$kg s$^{-1}$) increase in the dust production rates.  

\subsection{Nucleus}

Aperture photometry mixes light scattered from the nucleus with light scattered from near-nucleus dust.  The least model-dependent upper limit to the  nucleus is set by $V_{0.2}$, which corresponds to scattering cross-sections of 3.32$\pm$0.03 km$^2$ and 3.08$\pm$0.03 km$^2$ on October 14 and 28, respectively (Table \ref{photometry}).  A limit to the effective circular radius of the nucleus is computed from $r_n = (C_e/\pi)^{1/2}$, giving $r_n$ = 1.03 km on the former date and $r_n$ = 0.99 km on the latter.  The statistical uncertainties on these radii are $<$1\% but the actual uncertainties are much larger, possibly rising to several tens of percent, as a result of the unmeasured phase function and geometric albedo.   In any case, this measurement at best sets only an upper limit to the nucleus radius, which we take as $r_e \lesssim 1$ km.

\begin{figure}
\epsscale{0.90}
\begin{center}
\plotone{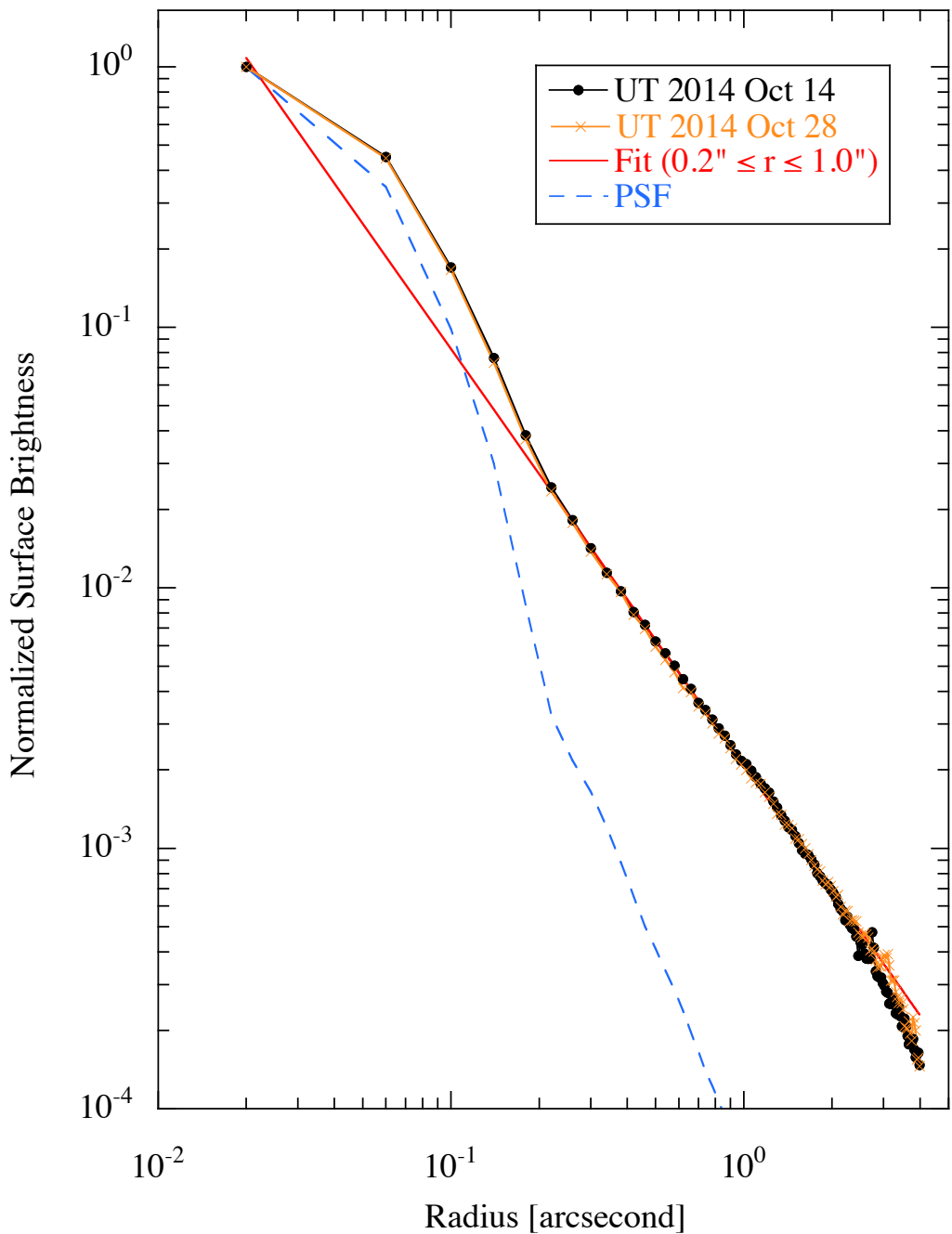}
\caption{Surface brightness profile of 313P measured on UT 2014 October 14 (black line and symbols) compared with a power-law fit to the 0.2\arcsec~$\le r \le$ 1.0\arcsec~radius range (red line).  The fit has an index $p$ = 1.64$\pm$0.01. The point spread function of the WFC3 camera is also shown for reference (blue dashed line).   All profiles are normalized to unity at the central pixel. \label{profile}
} 
\end{center} 
\end{figure}

To attempt to place a more stringent limit on the nucleus radius, we next determined the surface brightness profile from the HST data using  circular photometry annuli centered on the nucleus. Background subtraction was obtained from a larger, concentric annulus having inner and outer radii 6.0\arcsec~and 12.0\arcsec, respectively (Figure \ref{profile}).  The resulting profiles from October 14 and 28 are nearly indistinguishable at the scale of the figure but quite different from the normalized point spread function (PSF) of WFC3, as computed using the TinyTim modeling software (Krist et al.~2011).    For all radii $r \gtrsim$ 0.2\arcsec, the surface brightness of 313P exceeds that of the normalized PSF by an order of magnitude or more as a result of scattering by dust.

In the radius range 0.2\arcsec~$\le r \le$ 1.0\arcsec, the coma surface brightness follows a power law, $\Sigma(r) \propto r^{-p}$, with a least-squares fit index $p$ = 1.64$\pm$0.01 on both October 14 and 28.  This index is steeper than the canonical $p$ = 1, expected of a steady-state, isotropic coma expanding at constant speed, but close to the $p$ = 3/2 expected of an expanding coma in which the grains are accelerated by solar radiation pressure (Jewitt and Meech 1987).  We used the fit to $\Sigma(r)$ to extrapolate the surface brightness inwards to $r$ = 0, then integrated $\int_0^{0.2}2\pi r \Sigma(r) dr$ to estimate the brightness expected of the coma in the absence of a contribution from the nucleus.  We define $\mathcal{R}$ as the ratio of the cross-section of the nucleus, $C_n$, to the total cross-section of nucleus plus dust  in the central 0.2\arcsec, the latter given by $C_e$ from Table (\ref{photometry}). We find $\mathcal{R}$ = 0.25$\pm$0.02 on October 14 and  $\mathcal{R}$ = 0.28$\pm$0.02 on October 28.  With $C_e$ from Table (\ref{photometry}) we obtain $C_n$ = 0.83$\pm$0.07 km$^2$ and 0.86$\pm$0.06 km$^2$ on October 14 and 28, respectively.  These cross-sections correspond to equal-area spheres of radii $r_n$ = 0.51$\pm$0.02 km and 0.52$\pm$0.02 km, respectively.  The formal uncertainties on $r_n$, as well as the close agreement between the two values, suggest a pleasing accuracy which, however, is greater than realistically possible from our data.  We have no evidence, for example, that $\Sigma(r)$ follows the same (or any) power law down to the nucleus, the phase function is unmeasured and the value of the geometric albedo is little more than a guess.  Nevertheless, we can reasonably conclude from the photometry that the nucleus of 313P is a sub-kilometer body, with a best estimate of the radius being $r_n$ = 0.5 km.  

We also searched the HST images for evidence of companion nuclei, such as might be produced by fragmentation of the primary body.  No such companion nuclei were found.  To set limits to the brightness of any such objects, we placed artificial objects on the field and measured the faintest that could be reliably detected by eye.  The complexity of the surface brightness distribution  prevents us from asserting a single value of the upper limit to the size of companion nuclei.  However, over most of the field we could reliably detect objects with V $\le$ 27.8 (measured within a 0.2\arcsec~radius circle). This is $\sim$7 magnitudes fainter than the primary, giving an upper limit to the radius of about 25 m, assuming 0.05 geometric albedo.  Weaker limits (and larger limiting radii) apply within an arcsecond of the nucleus and in the brighter regions of the tail seen in Figure (\ref{images}).

\subsection{Spectrum}
\label{spectra}
The optical spectrum was measured on UT 2014 October 22, in order to search for emission features due to gas.  In the optical spectra of comets, the brightest accessible emission line is the bandhead of CN at 3888\AA.  Accordingly, we employed the  spectroscopic mode of LRIS with the 400/3400 grism, giving a dispersion of 1.07 \AA~pixel$^{-1}$ and a 1\arcsec~wide slit, yielding a resolution of $\sim$7\AA~FWHM near 3900\AA.  A total of 2700 s of spectral data were secured, divided into three separate integrations each of 900 s in order to facilitate the removal of cosmic rays.  We used images of field stars to measure the fraction of the light  passing through the 1.0\arcsec wide slit as $f_s$ = 0.4 (the instantaneous seeing for this measurement was 0.9\arcsec~FWHM).  Beyond securing a sufficient signal, the main problem in the search for gas is the imprint of telluric and solar absorptions on the asteroid spectrum.  In the past, we have used solar analogue stars to cancel these features, but rarely find these stars to offer a good match to the absorption line spectrum of the Sun.  Instead, we observed the nearby $V \sim$ 13.0 asteroid (458) Hercynia reasoning that, although its continuum slope may differ from that of 313P (since Hercynia is an S-type), the all-important fine-scale spectral structure should be exactly the same.  

\begin{figure}
\epsscale{0.95}
\begin{center}
\includegraphics[width=0.95\textwidth,angle=-0]{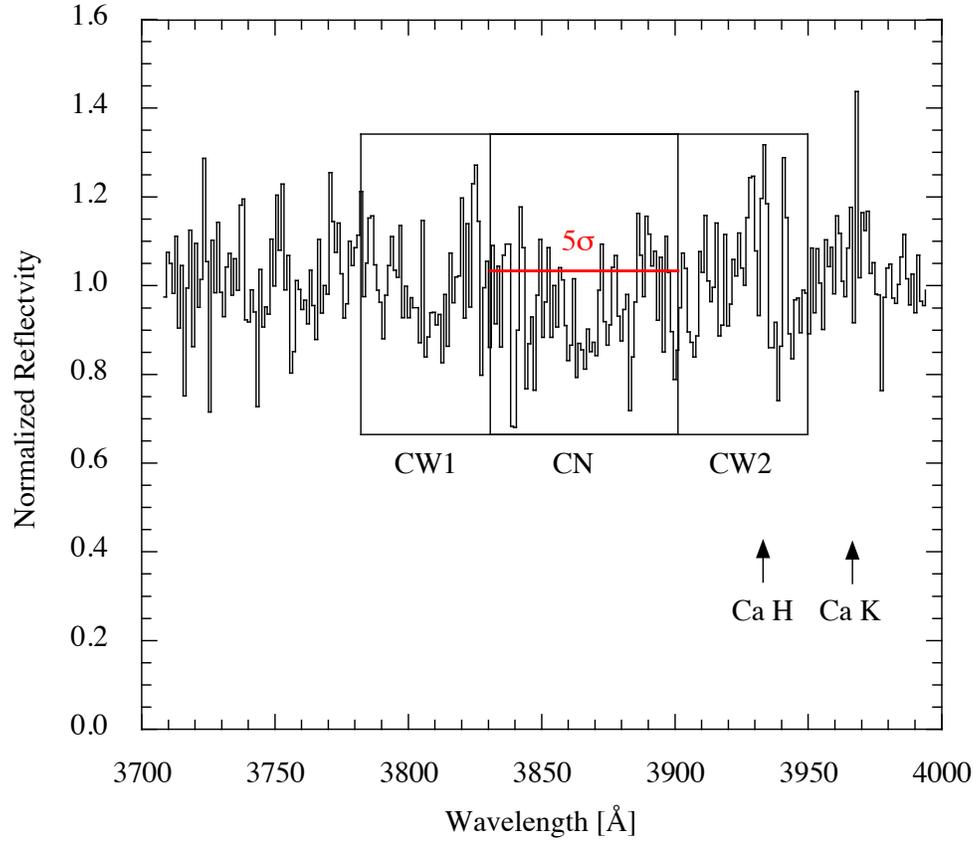}
\caption{Normalized reflection spectrum from UT 2014 October 22 showing the locations of the flanking continuum wavelengths, CW1 and CW2, and the band used to extract a limit to CN.  The wavelengths of Ca H and K lines are marked.  The horizontal red line marks the derived 5$\sigma$ upper limit to the CN band.   \label{spectrum}
} 
\end{center} 
\end{figure}

Figure (\ref{spectrum}) shows the reflection spectrum of 313P, extracted from a 2.7\arcsec~wide box around the object, with sky subtraction from a parallel region 5.4\arcsec~from 313P and divided by a similarly sky-subtracted spectrum of asteroid (458) Hercynia.  The resulting ratio has been further divided by a second-order least-squares fit polynomial in order to remove large scale spectral variations.

The magnitude of 313P measured at the time of the spectrum in Figure (\ref{spectrum}) was B = 20.39$\pm$0.02, corresponding to the flux density $f_{B}$ = 4.7$\times$10$^{-17}$ erg cm$^{-2}$ s$^{-1}$ \AA$^{-1}$ at the central wavelength 4500\AA.  Since the continuum color of 313P is nearly neutral with respect to the Sun, we estimate the continuum flux density at the wavelength of CN (3888\AA) from $f_{3888} = f_{B} f_s [f_{3888}/f_{4500}]_{\odot}$, where  $[f_{3888}/f_{4500}]_{\odot}$ = 0.52 is the ratio of the flux densities in the Sun at the wavelengths of CN and of the B-filter center (Arvesen et al.~1969).  Therefore, the continuum flux density at the CN band is 9.8$\times$10$^{-18}$ erg cm$^{-2}$ s$^{-1}$ \AA$^{-1}$.  Based on the standard deviation of the local continuum, the 5$\sigma$ limit to the flux density from CN is 9.8$\times$10$^{-19}$ erg cm$^{-2}$ s$^{-1}$ \AA$^{-1}$, corresponding to an average flux across the 70\AA~CN window of 6.9$\times$10$^{-17}$ erg cm$^{-2}$ s$^{-1}$

The 1.0\arcsec$\times$2.7\arcsec~slit samples only a fraction of the total area of the coma potentially occupied by CN molecules, requiring the use of a model to correct to the total population.  We integrated a Haser model over the above slit area for this purpose.  With g-factor, $g$ =    4.0$\times$10$^{-14}$ erg s$^{-1}$ molecule$^{-1}$ (Schleicher 2010) and an assumed outflow speed of 500 m s$^{-1}$, we obtain 5$\sigma$ limit to the CN production rate $Q_{CN} <$ 1.8$\times$10$^{23}$ s$^{-1}$.  In comets, the average ratio of CN to water production rates is $Q_{H_2O}/Q_{CN} \sim$ 360 (A'Hearn et al.~1995).  If applicable to 313P, this ratio would imply a 5$\sigma$ limiting production rate $Q_{H_2O} <$ 6$\times$10$^{25}$ s$^{-1}$, corresponding to 1.8 kg s$^{-1}$.   This low value is orders of magnitude smaller than found in typical near-Sun Jupiter family comets, but representative of rates deduced spectroscopically for other active asteroids (Jewitt 2012).  While the non-detection of CN is reliable, the meaning of the derived $Q_{H_2O}$ is open to question, since  $Q_{H_2O}/Q_{CN}$ may take different values in asteroids and comets.


In 313P, the derived limit to the production rate in gas ($<$1.8 kg s$^{-1}$) is formally consistent with the increase in the dust production rate ($\sim$0.2 to 0.4 kg s$^{-1}$) estimated from photometry in Section (\ref{dust}).  However, comparison of these rates to estimate the dust/gas ratio in 313P is complicated by our lack of knowledge of the temporal variability of the mass loss.  Gas molecules, traveling at $V \sim$ 500 m s$^{-1}$ leave the vicinity of the nucleus much more quickly than do dust particles (ejection speeds $V \ll$ 4 to 6 m s$^{-1}$).  Therefore, it is possible that we did not detect gas in 313P simply because it had dissipated in the days before the spectroscopic observation on October 22.

\section{DISCUSSION}
The central issue is to identify  the source of the activity observed in 313P.  Observations of other active asteroids have revealed a remarkably diverse set of mechanisms (Jewitt 2012).  Included are the expulsion of debris following asteroid-asteroid impact (c.f.~(596) Scheila, Bodewits et al.~2011, Ishiguro et al.~2011, Jewitt et al.~2011),  mass-shedding of particulate regolith material (311P/PanSTARRS, Jewitt et al.~2013, Hainaut et al. 2014, Moreno et al.~2014),  rotational break-up (P/2013 R3, Jewitt et al.~2014a),  thermal fracture ((3200) Phaethon, Li and Jewitt 2013) and the sublimation of exposed ice (133P/Elst-Pizarro, Hsieh et al.~2004, Jewitt et al.~2014b; 238P/Read, Hsieh et al.~2011).  

Even from limited data, we can begin to eliminate some of these mechanisms as likely drivers of the mass loss in 313P.  Impact should produce a nearly instantaneous rise to maximum light followed by steady fading as the debris recedes from the nucleus.  Our data are inconsistent with an impact origin because the photometry does not show the expected, steady fading (Table \ref{photometry}).   The heliocentric distance is too large, and the temperature too low, to suggest that thermal fracture and desiccation are significant dust sources in 313P.  We also reject rotational mass shedding, as observed in 311P/PanSTARRS (Jewitt et al.~2013, Hainaut et al. 2014, Moreno et al.~2014)  where dust is briefly released with zero initial velocity, producing synchrone-like dust tails.  In contrast, the dust emitted from 313P is better matched by protracted emission of particles (syndynes).   Furthermore, 313P shows no multiple nucleus structure or evidence for companions (with a limiting radius near 25 m), providing no evidence for break-up of the sort seen in P/2013 R3 (Jewitt et al.~2014a).   

The observed protracted ejection of material is consistent with the action of sublimation.  At $R$ = 2.4 AU, a dirty ice surface (Bond albedo 0.05) exposed at the subsolar point sublimates in equilibrium with sunlight at the specific rate $F_s$ = 5.6$\times$10$^{-5}$ kg m$^{-2}$ s$^{-1}$.  The area of an exposed ice patch needed to sustain mass loss at the rate $dM_d/dt$ is

\begin{equation}
\pi r_s^2 = \left(\frac{1}{f_{dg} F_s} \right) \frac{dM_d}{dt}
\end{equation}

\noindent where $r_s$ is the circular equivalent radius of the sublimating patch and $f_{dg}$ is the ratio of the dust to gas production rates.  We conservatively take $f_{dg}$ = 10 (the value in 2P/Encke is 10 $\le f_{dg} \le$ 30, Reach et al.~2000) and substitute $dM_d/dt$ = 0.2 to 0.4 kg s$^{-1}$ to find $r_s$ = 11 to 15 m.  Ice located away from the subsolar point would sublimate less rapidly but, even so, we obtain values of $r_s$ very small compared to the 500 m radius of the nucleus of 313P, with fractional areas corresponding to only $\sim$10$^{-3}$ of the surface of the asteroid.  Sublimation would naturally account for the protracted ejection shown by 313P, and might explain why 313P was only just discovered, since sublimation is expected to be strongest close to perihelion (UT 2014 August 28).   The rate of recession of a sublimating surface is $\sim F_s/\rho$.  Given bulk density $\rho$ = 1000 kg m$^{-3}$, the surface would recede at $\sim$560 \AA~s$^{-1}$.  In the $\sim$1 month period of observations described here, the surface would recede by a distance $\sim$10 cm.  The nature of the trigger needed to expose ice in the first place is unknown, but could be a disturbance by an impact or a local avalanche.

\begin{figure}
\epsscale{0.65}
\begin{center}
\includegraphics[width=0.95\textwidth,angle=0]{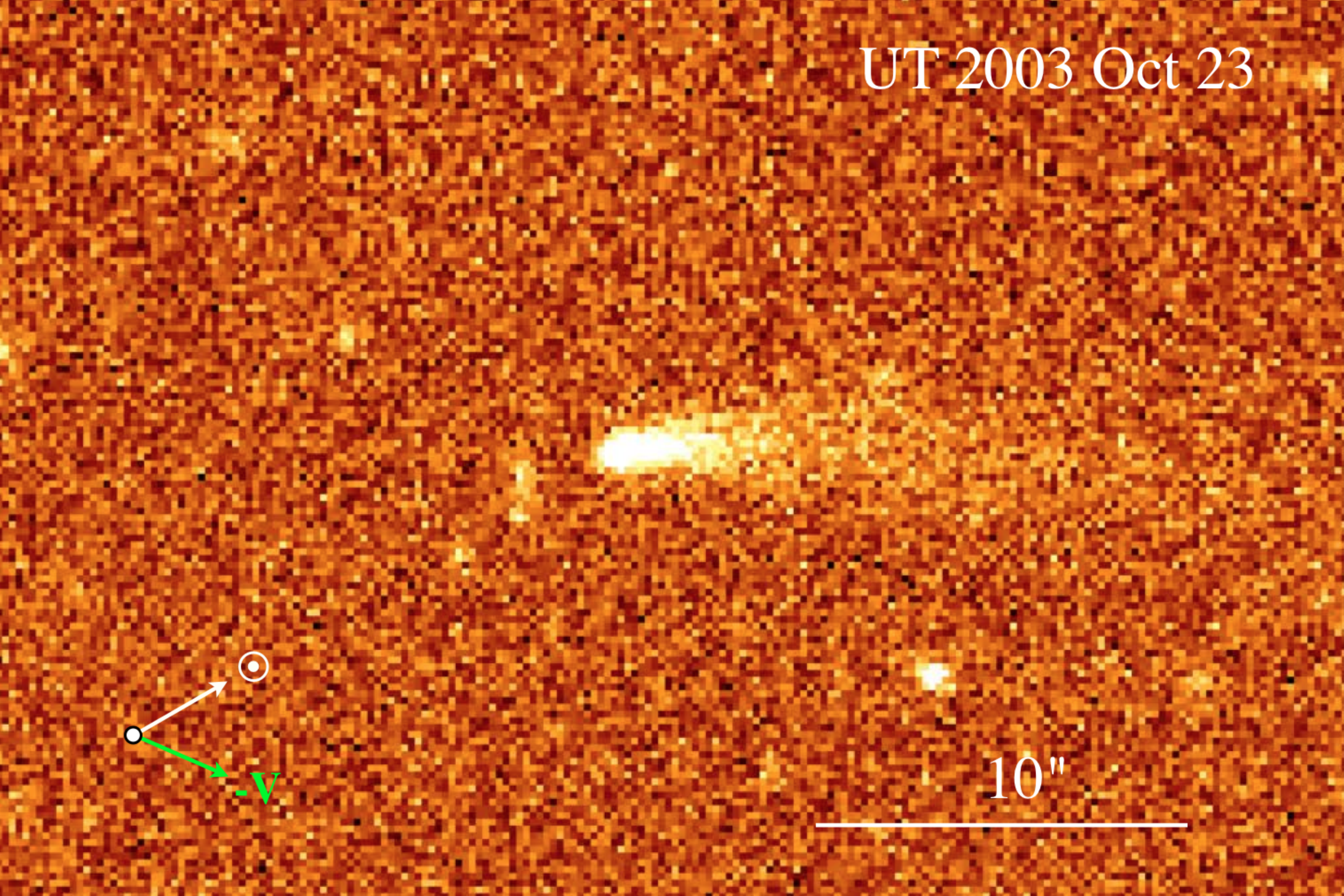}
\caption{Image of 313P recorded in archival Sloan Digital Sky Survey data from UT 2003 October 23.  This is a 54 s integration through a Sloan r filter.  The image has North to the top, East to the left. Direction arrows and a scale bar are shown for reference.  \label{2003oct23}
} 
\end{center} 
\end{figure}

Even more convincing evidence of a sublimation origin for the activity is the existence of archival data showing that 313P was active when near perihelion in 2003.  For example, observations from UT 2003 October 23 (Figure \ref{2003oct23}) show a prominent dust tail at heliocentric distance $R$ = 2.464 AU (perihelion was at 2.367 AU on UT 2003 June 20), essentially the same distance as in the new observations reported here (Table \ref{geometry}).  This puts 313P in the same category of active asteroids as 133P/Elst-Pizarro and 238P/Read, both of which exhibit protracted episodes of activity at more than one perihelion (Hsieh et al.~2004, 2011, 2013, Moreno et al.~2013).   Measurements from the archival data are underway and will be reported in a later paper.  The elongation of 313P at the previous perihelion (UT 2009 January 19) was less favorable for optical observations and we are aware of no reported detections.

We note that the orbit of 313P is very close to those of active asteroids 238P and P/2012 T1, and that the semimajor axis is close to that of 133P (Figure \ref{ae_plot}).  The evidence thus points to a group of active asteroids in the outer belt exhibiting similar photometric properties and, presumably, being driven by a common process, which we believe to be the sublimation of near-surface ice.

One difference between 313P and 133P is that, whereas the former displays a prominent fan-shaped dust tail, the latter is morphologically distinguished instead by a parallel-sided trail of more slowly-moving, older and much larger particles (up to at least centimeter size; Jewitt et al.~2014b).   Part of the difference in appearance between the two could be a result of the different viewing geometries, with 313P observed from $\sim$7\degr~above the orbital plane (Table \ref{geometry}) while the corresponding angle for observations of 133P is $\le$1.4\degr.  However, the complete absence of material appearing along the projected orbit  (c.f.~Figure \ref{models}) suggests that 313P is deficient in the very largest, most slowly-moving particles, or that they have yet to travel far enough from the nucleus to populate a distinct trail.  We conjecture that 313P is a ``young'' version of 133P, and is dominated by the continuing ejection of particles of modest size.  If so, we expect that its morphology will evolve to more closely resemble that of 133P as large particles drift slowly away from the nucleus to form a parallel-sided trail.  Meanwhile, small-to-moderate sized particles should disperse under the action of solar radiation pressure.  Additional observations are planned to monitor the development of 313P, to show whether or not it becomes more 133P-like, and to better define the progress of its activity.

Lastly, we remark on the likely connection between sublimation and rotational disruption.   Mass-loss torques on a sublimating icy body  can be orders of magnitude larger than radiation torques, resulting in very short timescales for changing the spin.  
The e-folding timescale for changing the spin of a body from the torque due to non-central mass-loss is (Jewitt 1997)

\begin{equation}
\tau_s \sim \frac{\omega \rho r_n^4}{V k_T dM/dt}
\label{torque}
\end{equation}

\noindent where $\omega$ is the initial angular frequency of the rotation, $\rho$ is the bulk density, $V$ and $dM/dt$ are the speed and rate at which mass is ejected and $k_T$ is the dimensionless moment arm (i.e.~the moment arm of the torque expressed in units of the radius, $r_n$).  The $r_n^4$ dependence in Equation (\ref{torque}) renders the small nucleus of 313P ($r_n$ = 0.5 km) highly susceptible to spin-up by outgassing torques and raises the possibility of future rotational break-up.


To estimate  $\tau_s$ from Equation (\ref{torque}), we note that the limiting values of $k_T$ are 0, corresponding to mass loss in a jet radial to the nucleus surface and 1, corresponding to a jet tangential to the surface.  Images of comets from spacecraft show that the jets are more nearly radial than tangential, meaning that $k_T \ll$ 1.  A simplistic model gives $k_T \sim$ 0.05 (Jewitt 1997), while measurements of spin-changes give 0.005 $\le k_T \le$ 0.04 for active Jupiter family comet P/Tempel 1 (Belton et al.~2011) and $k_T$ = 4$\times$10$^{-4}$ for P/Hartley 2 (Drahus et al.~2011).  We take 10$^{-3}$ $\le k_T \le$ 10$^{-2}$ as a compromise and adopt $\omega$ = 3.5$\times$10$^{-5}$ s$^{-1}$, corresponding to a nominal 5 hour rotation period, $\rho$ = 1000 kg m$^{-3}$ and $V$ = 500 m s$^{-1}$ as the speed of the outflowing gas (Biver et al.~1997).  Then, setting $dM/dt \lesssim$ 1 kg s$^{-1}$, the maximum value allowed by the spectroscopic non-detections of gas (Section \ref{spectra}), we find a sublimation-driven spin-up time  $\tau_s \gtrsim$ 10 to 100 yrs. This estimate should be increased to account for the fact that outgassing is likely to be sustained only over the portion of the orbit near perihelion.  Even so, this very short time (ranging from a few to a few tens of orbital periods) shows that spin-up by outgassing can efficiently change the spins of small active asteroids, even with outgassing rates less than 1 kg s$^{-1}$.  Rotation and sublimation working together have already been implicated in the prototype active asteroid 133P/Elst-Pizarro (Jewitt et al.~2014b) whose spin-up time ($r_n$ = 2.2$\pm$0.5 km; Hsieh et al.~2004, Jewitt et al.~2014b) is, however, much longer  (5500 $\le \tau_s \le$ 55,000 yrs under the same assumptions as listed above).   Asteroid (62412) 2000 SY178 is also known to be a rapid rotator (Sheppard and Trujillo 2014).  The tiny precursor  to disrupted asteroid P/2013 R3 (estimated radius $r_n \sim$ 350 m, Jewitt et al.~2014a) could have been quickly driven to breakup by outgassing torques, if ice were exposed on its surface.  Outgassing from 313P may yet drive the nucleus to rotational break-up.





\clearpage

\section{SUMMARY}

313P/Gibbs is a $\sim$0.5 km radius (geometric albedo 0.05 and C-type asteroid phase function assumed) body located in the outer asteroid belt. Observed at Hubble Space Telescope resolution ($\sim$84 km) it is a single body that emits dust at rates $\sim$0.2 to 0.4 kg s$^{-1}$.  Both aperture photometry (showing a scattering cross-section that increases with time) and dynamical analysis of the tail show that dust ejection is protracted (not impulsive).  No emission lines are spectroscopically detected, setting (indirect) limits to the production rate of water $<$1.8 kg s$^{-1}$.  

The cause of the activity in 313P cannot be definitively established from the available data. Nevertheless, some explanations are more plausible than others and some can be ruled-out.  Asteroid-asteroid impact is ruled-out because it should produce dust impulsively, followed by monotonic fading that is not observed.  There is no evidence for rotational break-up under the action of YORP or other torques, since the nucleus remains singular.  Another kind of rotational instability, so-called ``mass shedding'' (as seen in active asteroid 311P), produces synchronic tails which are absent in 313P/Gibbs.  Our preferred explanation is that equilibrium sublimation of dirty water ice from an exposed patch covering only a few hundred square meters ($\sim$10$^{-3}$ of the surface)  accounts for the measured dust production.  Sublimation naturally explains the protracted nature of the activity, and its recurrence near perihelion in both 2003 and 2014.  313P/Gibbs shows both physical and orbital similarities to several other suspected ice-sublimating bodies, consistent with a high abundance of ice in the outer asteroid belt near semimajor axis 3.1 AU.

%
%
%
%
%
%
%
%
%

\acknowledgments
We thank Pedro Lacerda for pointing out the 2003 archival data and the anonymous referee for comments. Based in part on observations made with the NASA/ESA \emph{Hubble Space Telescope,} with data obtained at the Space Telescope Science Institute (STSCI).  Support for program 13864  was provided by NASA through a grant from STSCI, operated by AURA, Inc., under contract NAS 5-26555.  We thank Linda Dressel, Alison Vick and other members of the STScI ground system team for their expert help.   We thank Joel Aycock and Greg Wirth for assistance at the  W.M. Keck Observatory, operated as a scientific partnership among Caltech, the University of California and NASA. The Observatory was made possible by the generous financial support of the W. M. Keck Foundation.  NP acknowledges funding by the Gemini-Conicyt Fund, allocated to project N\textsuperscript{\underline{o}} 32120036.

\clearpage

\clearpage

\begin{deluxetable}{lclccccccc}
\tablecaption{Observing Geometry 
\label{geometry}}
\tablewidth{0pt}
\tablehead{ \colhead{UT Date and Time} & DOY\tablenotemark{a}  & $\Delta T_p$\tablenotemark{b} & Tel\tablenotemark{c} & \colhead{$R$\tablenotemark{d}}  & \colhead{$\Delta$\tablenotemark{e}} & \colhead{$\alpha$\tablenotemark{f}}   & \colhead{$\theta_{\odot}$\tablenotemark{g}} &   \colhead{$\theta_{-v}$\tablenotemark{h}}  & \colhead{$\delta_{\oplus}$\tablenotemark{i}}   }
\startdata

2014 Oct 02 04:53 - 09:23 &  275 & 35 & D1.5  & 2.399 & 1.429 & 7.7 & 329.2 & 247.0 & 7.6 \\
2014 Oct 03 07:48 - 09:00 &  276 & 36 & D1.5  & 2.400 & 1.430 & 7.6 & 333.0 & 247.1 & 7.6 \\

2014 Oct 14 13:12 - 13:48 & 287 & 47 & H2.4 & 2.405 & 1.451 & 8.9 & 10.5 & 247.9 & 7.5 \\
2014 Oct 22 06:39 - 09:05  & 295 & 55 & K10 & 2.410 & 1.483 & 11.0 & 27.6 & 248.5 & 7.2\\
2014 Oct 23 09:10 - 09:33  & 296 & 56 & K10 &2.411 & 1.489 & 11.3 & 29.3 & 248.5 & 7.1  \\
2014 Oct 28 21:25 - 23:16 & 301 & 61 &  H2.4 & 2.415 & 1.522 & 13.0 & 37.1 & 248.8 & 6.8 \\

2014 Nov 06 04:17 - 04:40 & 310 & 70 &  D1.5 & 2.422& 1.464 & 15.5 & 45.4 & 249.2 & 6.1 \\
\enddata


\tablenotetext{a}{Day of Year, UT 2014 January 01 = 1}
\tablenotetext{b}{Number of days past perihelion}
\tablenotetext{c}{Telescope: D1.5 = ESO/Danish 1.5 m diameter telescope, La Silla, Chile; H2.4 = Hubble Space Telescope, 2.2 m diameter; K10 = Keck I 10 meter telescope, Mauna Kea, Hawaii}
\tablenotetext{d}{Heliocentric distance, in AU}
\tablenotetext{e}{Geocentric distance, in AU}
\tablenotetext{f}{Phase angle, in degrees}
\tablenotetext{g}{Position angle of the projected anti-Solar direction, in degrees}
\tablenotetext{h}{Position angle of the projected negative heliocentric velocity vector, in degrees}
\tablenotetext{i}{Angle of Earth above the orbital plane, in degrees}

\end{deluxetable}

\clearpage

\begin{deluxetable}{lccccr}
\tablecaption{Photometry
\label{photometry}}
\tablewidth{0pt}
\tablehead{
\colhead{UT Date}    & \colhead{$\Phi$\tablenotemark{a}   }& \colhead{Filter}     & \colhead{$m_{\lambda}\tablenotemark{b}$} & \colhead{$m_{\lambda}(1,1,0)$\tablenotemark{c}} & \colhead{$C_e [km^2]\tablenotemark{d}$} 
}

\startdata
October 02 & 6.0 & R & 19.32$\pm$0.02 & 16.09$\pm$0.02  & 7.13$\pm$0.14 \\\\

October 03 & 6.0 & R & 19.33$\pm$0.01 & 16.11$\pm$0.01  & 7.06$\pm$0.07 \\\\

October 14 & 0.2 & V & 20.62$\pm$0.01 & 17.30$\pm$0.01  & 3.32$\pm$0.03 \\
October 14 & 1.0 & V & 20.02$\pm$0.01 & 16.70$\pm$0.01  & 5.77$\pm$0.06 \\
October 14 & 4.0 & V & 19.60$\pm$0.01 & 16.28$\pm$0.01  & 8.50$\pm$0.09 \\
October 14 & 6.0 & V & 19.40$\pm$0.01 & 16.08$\pm$0.01  & 10.21$\pm$0.10 \\\\

October 22 & 6.0 & B & 20.39$\pm$0.02 & 16.94$\pm$0.02  & 8.97$\pm$0.18 \\
October 22 & 6.0 & V & 19.67$\pm$0.01 & 16.22$\pm$0.01  & 9.06$\pm$0.09 \\
October 22 & 6.0 & R & 19.31$\pm$0.01 & 15.86$\pm$0.01  & 8.81$\pm$0.09 \\\\

October 23 & 6.0 & B & 20.43$\pm$0.02 & 16.96$\pm$0.02  & 8.81$\pm$0.17 \\
October 23 & 6.0 & V & 19.71$\pm$0.02 & 16.24$\pm$0.02  & 8.73$\pm$0.18 \\
October 23 & 6.0 & R & 19.31$\pm$0.02 & 15.84$\pm$0.02  & 8.97$\pm$0.17 \\\\

October 28 & 0.2 & V & 20.97$\pm$0.01 & 17.38$\pm$0.01  & 3.08$\pm$0.03 \\
October 28 & 1.0 & V & 20.36$\pm$0.01 & 16.77$\pm$0.01  & 5.41$\pm$0.05 \\
October 28 & 4.0 & V & 19.81$\pm$0.01 & 16.22$\pm$0.01  & 8.97$\pm$0.09 \\
October 28 & 6.0 & V & 19.72$\pm$0.01 & 16.13$\pm$0.01  & 9.75$\pm$0.10 \\

November 06 & 6.0 & R & 19.65$\pm$0.07 & 16.05$\pm$0.07 & 7.39$\pm$0.51 \\
   
\enddata


\tablenotetext{a}{Projected angular radius of photometry aperture, in arcseconds}
\tablenotetext{b}{Apparent magnitude at the wavelength, $\lambda$, corresponding to the filter employed}
\tablenotetext{c}{Magnitude corrected to $R$ = $\Delta$ = 1 and $\alpha$ = 0\degr~by Equation (\ref{absolute})}
\tablenotetext{d}{Effective scattering cross-section computed from Equation (\ref{area})}

\end{deluxetable}

\clearpage


\clearpage

\clearpage

\clearpage

\clearpage

\clearpage

\clearpage

\clearpage

%
%

%
%
%
%
%
%

\end{document}